# Equivalence between the pair correlation functions of primes and the four-point correlation functions of spins in a two-dimensional Ising model with randomly distributed competing interactions

Zhidong Zhang

Shenyang National Laboratory for Materials Science, Institute of Metal Research,

Chinese Academy of Sciences, 72 Wenhua Road, Shenyang, 110016, P.R. China

## Abstract

In this work, we prove the equivalence between the pair correlation functions of primes, and the four-point correlation function of spins in a two-dimensional (2D) Ising model $M_{FI+SGI}^{2D}$, with a mixture of ferromagnetic and randomly distributed competing interactions. At first, we prove that the four-point connected correlation function of spins in the competing 2D Ising model is larger than zero in the temperature range from zero ($T$ = 0) to infinite ($T \to \infty$), because of nonlinear terms of spin operators. Second, we prove that the four-point connected correlation function of spins is equivalent to the pair correlation function of energy levels in the competing 2D Ising model. Third, we prove that the pair correlation function of the energy levels of the competing 2D Ising model is equivalent to the pair correlation function of nontrivial zeros of the Dirichlet function $L(s, \chi_k)$ (including the Riemann zeta function $\zeta(s)$). Fourth, we prove that the pair correlation function between the nontrivial zeros of the Dirichlet function $L(s, \chi_k)$ is equivalent to the pair correlation function of primes $p_n$ and $p_{n+\Delta n} = p_n + q$ with $\Delta n = 1, 2, \dots, \infty$ for every even $q$. In a conclusion, we have proven that the pair correlation function $\langle p_n p_{n+\Delta n} \rangle$ of the primes $p_n$ and $p_{n+\Delta n} = p_n + q$ for every even $q$ is larger than zero.

The corresponding author: Z.D. Zhang,

Tel: 86-24-23971859,

Fax: 86-24-23891320,

e-mail address: zdzhang@imr.ac.cn

## 1.Introduction

The de Polignac conjecture (including the twin prime conjecture as its particular case) is one of the most fundamental problems in mathematics [16,30], which is connected to many problems in mathematics and physics. The de Polignac conjecture states that there are infinitely many pairs of primes with a difference of an even number $q$ (=2, 4,…, $X$).

Let $p_n$ denote the $n$-th prime. The twin prime conjecture states that $\lim_{n\to\infty} inf(p_{n+1} - p_n) = 2$. In an equivalent expression, it states that the pair correlation function of twin primes is larger than zero as $n \to \infty$. Several approaches have attempted to solve the problem. 1) To search twin primes as large as possible by computing projects, below a large natural number $n$. However, the computational complexity increases rapidly as $n$ increases. So, it is too far for proving/disproving the twin prime conjecture, since it requires $n \to \infty$ out of capability of any computers. 2) To prove weaker conjectures. Along this streamline, Goldston et al. proved $\lim_{n\to\infty} inf \frac{p_{n+1}-p_n}{logp_n} = 0$ [25] and subsequently $\lim_{n\to\infty} inf \frac{(p_{n+1}-p_n)}{\sqrt{logp_n}(loglogp_n)^2} < \infty$ [26]. A significant progress was made by Y.T. Zhang [72], who proved $\lim_{n\to\infty} inf(p_{n+1} - p_n) < 7 \times 10^7$. Maynard [48] improved the above bound to 600, while the Polymath Research Project [56,57] subsequently improved this to 246. It is shown that there are infinitely many pairs of primes with some finite gap. However, a proof of the twin prime conjecture with a difference ending in 2 seems to be out of reach by this approach. 3) To establish the connection with other conjectures in mathematics with the same level of difficulties. For instance, under the generalized

Elliott–Halberstam conjecture [19], the best known bound is 6 [23,56]. The two-point correlation function for the nontrivial zeros of Dirichlet $L$-functions at a height $E$ on the critical line was calculated [8] heuristically using a generalization of the Hardy–Littlewood conjecture [30] for pairs of primes in arithmetic progression. 4) To connect the distribution of twin primes with physical systems. It is known that the density of the twin primes decreases with increasing $n$, and it is to prove that the correlation function of twin primes is larger than zero as $n \to \infty$. The main obstacle of the path is to find an appropriate spin model to satisfy the condition that the four-point correlation function of spins is equivalent to the pair correlation function of primes.

On the other hand, the Ising model is one of the most fundamental models in physics [36,38,39,54], which is associated with many problems in physics, mathematics and computer science. The study of the mathematical structures of the Ising model is helpful for a better understanding of other mathematical problems. The present author has solved two fundamental problems related with Ising models: the exact solution of the ferromagnetic three-dimensional (3D) Ising model [65,73,78,81,82] and the computational complexity of the spin-glass 3D Ising model [74,75,77,79]. In a preceding paper (defined as I [76]) the author proved the equivalence between the zero distributions of the Dirichlet function $L(s, \chi_k)$ (and the Riemann zeta function $\zeta(s)$) [11,58] and the partition function of a two-dimensional (2D) Ising model $M_{FI+SGI}^{2D}$ with randomly distributed competing interactions. By proofs of four theorems (Equivalence Theorem, Real Eigenvalues Theorem, Hilbert-Pólya Space Theorem, Unit Circle - Critical Line Theorem) [76], I proved the closure of the nontrivial zero distribution of the Dirichlet function $L(s, \chi_k)$ (including the Riemann zeta function $\zeta(s)$). Because the distribution of primes is closely related with the nontrivial zero distribution of the Dirichlet function $L(s, \chi_k)$, I believe that the competing 2D Ising model $M^{2D}_{FI+SGI}$ can play an important role also for solving the twin primes conjecture.

The aim of this work is to prove the equivalence between the pair correlation functions of primes, nontrivial zeros of the Dirichlet function $L(s, \chi_k)$ and the four-point correlation function of spins in the competing 2D Ising model $M^{2D}_{FI+SGI}$. In section 2, we review briefly current status in literatures of studies on the pair correlation function of primes based on some conjectures. In section 3, we first prove that the four-point connected correlation function of spins in the competing 2D Ising model is larger than zero in the temperature range from zero ($T$ = 0) to infinite ($T \to \infty$), because of nonlinear terms of spin operators. Then, we prove that the four-point connected correlation function of spins is equivalent to the pair correlation function of energy levels in the competing 2D Ising model. Next, we prove that the pair correlation function of the energy levels in the model $M^{2D}_{FI+SGI}$ is equivalent to the pair correlation function of nontrivial zeros of the Dirichlet function $L(s, \chi_k)$ (including the Riemann zeta function $\zeta(s)$). Furthermore, we prove that the pair correlation function between the nontrivial zeros of the Dirichlet function $L(s, \chi_k)$ is equivalent to the correlation function between a pair of primes $p_n$ and $p_{n+\Delta n} = p_n + q$ with $\Delta n = 1, 2, \dots, \infty$ for every even $q$. Namely, we have proven that the pair correlation function $\langle p_n p_{n+\Delta n} \rangle$ of the primes $p_n$ and $p_{n+\Delta n} = p_n + q$ for every even $q$ is larger than zero.

## 2. Notation and current status of studies on the pair correlation function of primes

*Notation.*

$p,\ p_n,\ p_{n+\Delta n}$: a prime number.

$q$, $X$, $h$, $h_1$, $h_2$, $h_3$: an even number.

$n,\ n'$, $N$, $N'$, $a$: an integer number.

$x$: a large number.

ε, $A$: a positive number.

$P_q(n)$: the number of pairs less than $n$

$C_2$: the twin prime constant.

$L(s,\chi_k)$: the Dirichlet function.

$\zeta(s)$: the Riemann zeta function.

$\mu(n)$: the Möbius function.

$\Lambda(n)$: the von Mangoldt function.

$P(\tilde{J}_2)$: the Gaussian function.

$\mathfrak{S}(h)$: a singular series.

$\chi_k(n)$: a primitive Dirichlet character.

$k,\ h'$: an integer modulo of the Dirichlet character.

$E$, $E_m$, $\epsilon,\ \epsilon_1,\ \epsilon_2$: the imaginary part of zeros of the Dirichlet function.

$d(E),\ \bar{d}(E),\ d^{osc}(E)$: the density of zeros of the Dirichlet function.

$R_2(\epsilon_1, \epsilon_2)$, $R_2^{osc}(\epsilon_1, \epsilon_2)$: the two-point correlation function of zeros of the Dirichlet function.

$\boldsymbol{H}$: the Hamiltonian of the spin system.

$s_{i,\mathrm{j}}$, $s_{i\prime,j\prime}$, $s_{u,\mathrm{v}}$, $s_{u\prime,v\prime}$: the Ising spin.

*i*, *j*, *u*, *v*, *t*, *l*, *m*: the lattice site.

$J_1$, $\tilde{J}_2$: the interaction between spins.

$K_1$, $K_1^*$, $\widetilde{K}_2$: the reduced interaction with respect to temperature.

$k_B$: the Boltzmann constant.

$T$: temperature.

Z, $\bar{Z}_\alpha$: the partition function.

α, *R*: a replica of the spin-glass system.

$\boldsymbol{R}$, $\boldsymbol{I}$, $\boldsymbol{O}$, $\boldsymbol{H}$: an operator.

$\langle \boldsymbol{O} \rangle_\alpha$: the thermal average of the operator $\boldsymbol{O}$ for a fixed replica α.

$\overline{\langle \boldsymbol{O} \rangle_\alpha}$: the disorder average of the operator $\boldsymbol{O}$ for a fixed replica α.

$\langle s_{i,j} s_{i\prime,j\prime} \rangle_\alpha$, $\overline{\langle s_{i,j} s_{i\prime,j\prime} \rangle_\alpha^2}$: the pair correlation function of spins.

$\langle s_{i,j} s_{i\prime,j\prime} s_{u,v} s_{u\prime,v\prime} \rangle_\alpha$, $\overline{\langle s_{i,j} s_{i\prime,j\prime} s_{u,v} s_{u\prime,v\prime} \rangle_\alpha}$: the four-point correlation function of spins.

$\overline{\Delta_\alpha}$, $\overline{\Delta'_\alpha}$: the four-point connected correlation function of spins.

$\gamma_{2j}$, $E_n$, $E_j$, $E_{j+\Delta j}$: an energy eigenvalue or an energy level.

$\omega_\alpha$: an angle in a hyperbolic triangle in the 2D Poincaré disk model.

$\Delta j$, $\Delta J$: the number of the gaps between energy levels.

$\langle E_j E_{j+\Delta j} \rangle_\alpha$, $\overline{\langle E_j E_{j+\Delta j} \rangle_\alpha}$, $\langle E_j E_{j+\Delta j} \rangle$: the correlation function of energy levels.

$L_s,\ L_{s+\Delta s}$: the nontrivial zeros of the Dirichlet function.

$\Delta s,\ \Delta S$: the number of the intervals between nontrivial zeros of the Dirichlet function.

$\langle L_s L_{s+\Delta s}\rangle$: the pair correlation function of nontrivial zeros of the Dirichlet function.

$\Delta n,\ \Delta N$: the number of the intervals between a pair of primes

$\langle p_n p_{n+\Delta n}\rangle$: the pair correlation function of the primes.

$\alpha(h)$: the density of prime pairs.

In this section, we review briefly the current status of studies on the pair correlation function of primes based on some conjectures, such as Hardy-Littlewood conjecture, Elliott–Halberstam conjecture, generalized Riemann hypothesis.

### 2.1 Hardy-Littlewood conjecture

Hardy and Littlewood investigated characters of pairs of primes [28-30,67]. The Hardy–Littlewood conjecture for pairs of primes is described as follows [30]:

**Hardy-Littlewood conjecture [30]:** There are infinitely many prime pairs

$$\omega, \omega' = \omega + q$$

for every even $q$ (= 2, 4,…, $X$). If $P_q(n)$ is the number of pairs less than $n$, then

$$P_q(n) \sim 2\mathrm{C}_2 \frac{n}{(\log n)^2} \prod_{\mathrm{p}} \left(\frac{p-1}{p-2}\right) \tag{1}$$

where $C_2$ is the twin prime constant as follows:

$$C_2 = \prod_{\omega=3}^{\infty} \left(1 - \frac{1}{(\omega-1)^2}\right) \approx 0.6601618.$$

(2)

and $p$ is an odd prime divisor of $q$. This conjecture is Conjecture B in [30], which is based on Hypothesis R with the Dirichlet functions $L(s,\chi_k)$.

The Dirichlet series are defined for the reciprocal of the Riemann zeta function $\zeta(s)$ [11,67]:

$$\frac{1}{\zeta(s)} = \prod_{p=2,3,5,7\ldots}\left(1-\frac{1}{p^s}\right) = \sum_{n=1}^{\infty}\frac{\mu(n)}{n^s} \qquad (Re(s)>1)$$

(3)

where $\mu(n)$ is the Möbius function.

The Dirichlet $L$-functions are defined as follows [67]:

$$L(s,\chi_k) = \sum_{n=1}^{\infty}\frac{\chi_k(n)}{n^s} = \prod_{p}\left(1-\frac{\chi_k(p)}{p^s}\right)^{-1}$$

(4)

for $\mathrm{Re(s)} > 1$ and by analytic continuation elsewhere, where $\chi_k(n)$ is a primitive Dirichlet character modulo an integer $k$ (= 1, 2, …, $h'$).

To investigate the correlation function between prime pairs, one usually denotes the von Mangoldt function $\Lambda$(n) [8,12,29,30]:

$$\Lambda(n) = \begin{cases} logp, & if\ \ n = p^k\ \ for\ \ some\ \ prime\ \ p \\ 0 & otherwise \end{cases}$$

(5)

Note that $\Lambda$(n) is defined for the condition $n = p^k$, in conjunction with the Dirichlet function $L(s,\chi_k)$.

Bogomolny and Leboeuf [9], Bogomolny and Keating [7,8] showed that the two-point correlation function of the nontrivial zeros of a Dirichlet function $L(s,\chi_k)$ tends in the universal limit (i.e. infinitely high up the critical line) to Gaussian unitary ensemble (GUE)/circular unitary ensembles (CUE) statistics of random matrix theory. Using a generalization of the Hardy–Littlewood conjecture for pairs of primes in arithmetic progression, the result [8] matched the conjectured random-matrix form in the limit as $E \to \infty$ and includes finite-$E$ corrections. Furthermore, the two-point correlation function result for the Riemann zeta function ζ(s) is essentially equivalent to a prime correlation formula that follows from the Hardy-Littlewood conjecture [24,40,51].

Furthermore, Keating and Smith [41] established, via a heuristic Fourier inversion calculation, that the Hardy–Littlewood twin-prime conjecture is equivalent to an asymptotic formula for the two-point correlation function of Riemann zeros at a height $E$ on the critical line and that the reverse implication also holds for this equivalence. Assuming the Riemann hypothesis, they calculated the pair correlation function in the sum of the diagonal terms $R_2^{diag}(\epsilon)$ and the off-diagonal terms $R_2^{off}(\epsilon)$.

### 2.2 Elliott–Halberstam conjecture

Elliott and Halberstam [19] proposed a conjecture described as follows:

**Elliott–Halberstam conjecture [19]:** Let $\pi(x,q,h)$ denote the number of primes not exceeding x that are congruent to h modulo q. Given any $A > 0$, we have:

$$\sum_{q \le X} \varphi(q) \max_{\substack{1 \le q \le h \\ (h,q)=1}} \left\{ \pi(x,q,h) - \frac{lix}{\varphi(q)} \right\}^2 \ll \frac{x^2}{log^A x} \tag{6}$$

provided:

$$\mathrm{X} \leq \frac{x^{\frac{1}{2}}}{log^{B}x}, \qquad B = B(A) > 0.$$

(7)

Assuming the Elliott-Halberstam conjecture, Goldston, Pintz and Yildirim [25] proved that there are infinitely often primes differing by 16 or less.

$$\lim_{n\to\infty} inf(p_{n+1} - p_n) \leq 16$$

(8)

Pintz [55] showed that assuming similar conditions not just for the primes but for functions involving both the primes and the Liouville function, one can assure not only the infinitude of twin primes but also the existence of arbitrarily long arithmetic progressions in the sequence of twin primes. Under the generalized Elliott-Halberstam conjecture, Polymath [56] showed the stronger statement that for any admissible triple ($h_1$, $h_2$, $h_3$), there are infinitely many $n$ for which at least two of $n + h_1$, $n + h_2$, $n + h_3$ are prime, and obtain a related disjunction asserting that either the twin prime conjecture holds. Meanwhile, Polymath [56] also modified the parity problem argument to show that the $H_1 \leq 6$ bound is the best possible that one can obtain from purely sieve-theoretic considerations. Under the generalized Elliott–Halberstam conjecture [19], the best known bound was 6 obtained also in [23].

Murty and Vatwani [53] formulated a conjecture regarding the equidistribution of the Möbius function over shifted primes in arithmetic progressions, in conjunction with the Elliott–Halberstam conjecture. This conjecture for a fixed even integer $h$ can resolve the parity barrier and produce infinitely many primes $p$ such that $p + h$ is also prime. It is conjectured that [53]:

$$\sum_{n\leq x} \Lambda(n)\Lambda(n+h) \sim \mathfrak{S}(\mathrm{h})x$$

(9)

where $\mathfrak{S}(h)$ is the singular series defined as

$$\mathfrak{S}(h) = \prod_{n|h}\left(1+\frac{1}{p-1}\right)\prod_{n\nmid h}\left(1-\frac{1}{(p-1)^2}\right)$$

(10)

If $h$ = 2, this gives an asymptotic formula for the number of twin primes.

### 2.3 Generalized Riemann Hypothesis

**Generalized Riemann Hypothesis [59,69]:** All nontrivial zeros of the Dirichlet function $L(s,\chi_k)$ lie on the critical line with real part $\frac{1}{2}$.

Heath-Brown [31] determined gaps between primes, and the pair correlation of zeros of the Riemann zeta function $\zeta$(s). Goldston [22] assumed the Riemann hypothesis, and examined how well the function $S(E)$ can be approximated by a Dirichlet polynomial in the $L^2$ norm, which is the error term in the formula for the number of nontrivial zeros of the Riemann zeta function $\zeta$(s) above the real axis and up to height $E$ in the complex plane. Under the generalized Riemann hypothesis, the statistical properties of these zeros are interesting as $E \to \infty$. The explicit formula for the density of zeros of any Dirichlet function $L(s,\chi_k)$ has the form [8]:

$$d(E) = \sum_{n} \delta(E-E_n) = \bar{d}(E) + d^{osc}(E)$$

(11)

It contains the smooth part $\bar{d}(E)$ and the oscillating part $d^{osc}(E)$. The formal expression for the two-point correlation function $R_2(\epsilon_1,\epsilon_2)$ consist of the smooth part $\bar{d}^2(E)$ and the oscillating part $R_2^{osc}(\epsilon_1, \epsilon_2)$. As all these additional factors tend to 1 when $\epsilon \to 0$ the universal behavior of the two-point function for the Dirichlet zeros is the same as for the GUE/CUE.

Bui, Keating and Smith [10] established the equivalence of conjectures concerning the pair correlation function of the nontrivial zeros of the Dirichlet function $L(s, \chi_k)$ in the Selberg class [60,61] and the variances of sums of a related class of arithmetic functions over primes in short intervals. This extended the results of Goldston and Montgomery [24] and Montgomery and Soundararajan [52] for the Riemann zeta function $\zeta$(s) to other $L$-functions in the Selberg class [60,61]. The approach was based on the statistics of the zeros.

## 3. Pair correlation functions of primes, zeros of the Dirichlet function $L(\mathbf{s}, \boldsymbol{\chi}_{\mathbf{k}})$ and four-point correlation functions of spins in the competing 2D Ising model $\boldsymbol{M}_{\boldsymbol{FI+SGI}}^{\boldsymbol{2D}}$

In this section, we shall investigate the behaviors of the pair correlation function of primes $p_n$ with changing *n* and pay a special attention on the limit $n \to \infty$. This will be done by the following steps: 1) To study the four-point correlation function of spins in a 2D Ising model with randomly distributed competing ferromagnetic/antiferromagnetic interaction. 2) To set up the connections between the four-point correlation function of spins and the pair correlation function of primes.

### 3.1 Four-point correlation functions of spins in the competing 2D Ising model $\mathbf{M}_{\mathbf{FI+SGI}}^{\mathbf{2D}}$

**Theorem 1 (Positive Four-Point Spin Correlation Theorem).** The four-point connected correlation function in the competing 2D Ising model $M_{FI+SGI}^{2D}$ is larger than zero in the temperature range from zero ($T$ = 0) to infinite ($T \to \infty$), because of nonlinear terms of spin operators.

**Proof of Theorem 1.** The Hamiltonian of the competing 2D Ising model with the nearest neighboring interactions, $M_{FI+SGI}^{2D}$, is written as:

$$H = -\sum_{<i,j>}^{l,m} \left[ J_1 s_{i,j} s_{i+1,j} + \tilde{J}_2 s_{i,j} s_{i,j+1} \right] \tag{12}$$

Here the competing 2D Ising model $M_{FI+SGI}^{2D}$ is the same as Definition 1 in I [76]. Every Ising spin with $s_{i,j} = \pm 1$ with the nearest neighboring interactions is located on a 2D lattice with the lattice size *lm*. The interaction $J_1 > 0$ along the horizontal direction is ferromagnetic, while the one $\tilde{J}_2$ along the vertical direction is a randomly distributed competing ferromagnetic/antiferromagnetic interaction as in a spin-glass system [5,18,50]. The distribution $P(\tilde{J}_2)$ of values $\tilde{J}_2$ is taken to be a Gaussian function with a non-zero mean $J_0 \neq 0$ and a variance $\hat{J}_2^2$ [76]. Namely, the interaction $\tilde{J}_2$ varies bond by bond randomly with different signs in range of [-$J_2$, $J_2$] with $J_2 \geq 0$. The interaction $J_2$ varies from zero to infinite, to fit with the unbounded Gaussian distribution. Note that the interaction $\tilde{J}_2$ in the vertical direction may be with some arbitrary but fixed period (for instance all the chains with the same distribution), so that the present model can be viewed as the random layered frustration model [33,34,47,70]. As demonstrated in [76], the Griffiths variable transformation [27,83] connects the spin-1 and spin-1/2 Ising models, which verifies the relation between the Möbius function and the spin-1/2 Ising model via the Potts model [71]. Thus the competing 2D Ising model can be utilized to imitate the Möbius function $\mu$(n).

Because of the randomness of the interaction $\tilde{J}_2$, the transfer matrix has the character of a random matrix [17,62,68], and behaves as the GUE [42,49]. By pretending that the imaginary parts $E_m$ of the Riemann zeros are eigenvalues of a quantum Hamiltonian whose corresponding classical trajectories are chaotic and without time-reversal symmetry, Berry [3] obtained the mean square difference between the actual and average numbers of Riemann zeros near the *x*-th zero in an interval. Berry and Keating [4] gave the Riemann zeros and eigenvalue asymptotics. Forrester et al. [20] studied the distributions of energy eigenvalues of GUE and the zeros of the Riemann zeta function $\zeta(s)$, described by nonlinear equations. Bogomolny [6] discussed statistical properties of quantum eigenvalues for chaotic systems based on semiclassical trace formulas, and considered the statistics of the zeros of the Riemann zeta function $\zeta(s)$. Conrey and Snaith [14] investigated correlations of eigenvalues of a unitary matrix (and its conjugate transpose) and Riemann zeros. Languasco et al. [45] obtained a quantitative version of the Goldston-Montgomery theorem about the equivalence between the asymptotic behaviors of the mean-square of primes in short intervals, and of the pair correlation function of the nontrivial zeros of the Riemann zeta function $\zeta(s)$. For more details of the relation between the random matrix and the Riemann zeta function $\zeta(s)$, readers refer also to [1,7,21,35].

Each Dirichlet function $L(s,\chi_k)$ for a different integer $k$ would have a different Hamiltonian $\boldsymbol{H}$ for each replica α (= 1, 2,…, $R$). The partition function Z of the 2D Ising model $M_{FI+SGI}^{2D}$ can be calculated from the product of the partition functions $\bar{Z}_{\alpha}$ for all fixed replicas (α = 1, 2,…, $R$), $\mathrm{Z} = \prod_{\alpha=1}^{R} \bar{Z}_{\alpha}$ [76]. The zeros of the partition function $\bar{Z}_{\alpha}$ of the 2D Ising model $M_{FI+SGI}^{2D}$ for each replica are equivalent to the nontrivial zeros of the Riemann zeta function $\zeta(s)$. We have $\bar{Z}_{\alpha} = 0 \Leftrightarrow \zeta(s) = 0$. Meanwhile, the zeros of the partition function Z of the 2D Ising model $M_{FI+SGI}^{2D}$ are equivalent to the nontrivial zeros of the Dirichlet function $L(s,\chi_k)$.

Following the procedure developed by Onsager [54] and Kaufman [38], we derived the eigenvalues for our present model $M_{FI+SGI}^{2D}$ [76],

$$cosh\gamma_{2j} = cosh2K_1^{*}cosh2\widetilde{K}_2 - sinh2K_1^{*}sinh2\widetilde{K}_2cos\omega_{\alpha} \tag{13}$$

where $\omega_{\alpha} = \frac{2j\pi}{n}$. As usual, it is convenient to introduce variables $K_1 \equiv J_1/(k_BT)$ and $\widetilde{K}_2 \equiv \tilde{J}_2/(k_BT)$ instead of interactions $J_1$ and $\tilde{J}_2$, where $k_B$ is the Boltzmann constant and $T$ is temperature. The Kramers-Wannier relation gives the definition of $K_1^{*}$ in its dual lattice [44]: $K_1^{*} = -\frac{1}{2}ln(tanhK_1)$. The angle $\omega_{\alpha}$ with the subscript α emphasizes that it would differ for different replicas with different Gaussian distributions of $\widetilde{K}_2$. Because $\widetilde{K}_2$ is random, the transfer matrix no longer commutes with translations along the random direction. The introduction of the time average together with the application of Jordan algebra solves the non-commutative problems (see Theorem 4 in [82]). Since there are no nonlinear terms in the transfer matrices, no topological Lorentz transformation is needed for deriving the eigenvalues of the present model. The geometric relationships are those for a hyperbolic triangle [38,54], as represented in the 2D Poincaré disk model (see Figure 4 in [54]).

The random Hamiltonian (1) requires the double averaging procedure for spin-glass systems [2]. For any observable $\boldsymbol{O}$ being an arbitrary function of the spin variables $s_{i,j}$ and the interactions $J_1$ and $\tilde{J}_2$, one first computes the thermal average $\langle \boldsymbol{O} \rangle_\alpha$ using the Boltzmann weight at temperature $T$ for a fixed replica α. The thermal average of operators $\boldsymbol{O}$ for a fixed replica α in the competing 2D Ising model can be defined as:

$$\langle \boldsymbol{O} \rangle_\alpha = \sum_{\{s_{i,j}\}} \left\{ \frac{\boldsymbol{O} \cdot exp[-\beta H_\alpha]}{\bar{Z}_\alpha} \right\} \tag{14}$$

with the partition functions $\bar{Z}_\alpha = \sum_{\{s_{i,j}\}} exp[-\beta H_\alpha]$ for the fixed replica α. Afterwards, the disorder average is taken over the random interaction distribution $P(\tilde{J}_2)$,

$$\overline{\langle \boldsymbol{O} \rangle_\alpha} = \int \sum_{<ij>} dJ_{ij}\, P(J_{ij})\, \langle \boldsymbol{O} \rangle_\alpha \tag{15}$$

The pair correlation function of spins is defined as the average of the product of the spins at two sites [39]. The correlation function for two given lattice sites states what is the probability that the spins at the two sites are the same. The thermal average for the pair correlation function of spins for a fixed replica α in the competing 2D Ising model can be written as:

$$\langle s_{i,j} s_{i',j'} \rangle_\alpha = \sum_{\{s_{i,j}\}} \left\{ \frac{s_{i,j} s_{i',j'} \cdot exp[-\beta H_\alpha]}{\bar{Z}_\alpha} \right\}$$

(16)

It is noticed that the disorder average induces a non-dynamical gauge symmetry [2]. The Hamiltonian (1) is invariant under the gauge transformation $s_{i,\mathrm{j}} \to \varepsilon_j s_{i,\mathrm{j}}$ and $\tilde{J}_2 \to \varepsilon_j \varepsilon_{j+1} \tilde{J}_2$ with a random sign per site $\varepsilon_j = \pm 1$. Since the gauge-transformed couplings $\varepsilon_j \varepsilon_{j+1} \tilde{J}_2$ are just as probable as the original ones, the quenched mean value of $\overline{\langle \boldsymbol{O}\{s_{\iota,J}\} \rangle_\alpha}$ is identical to that of its gauge average $\frac{1}{2^{lm}} \sum_{\{\varepsilon_j = \pm 1\}} \overline{\langle \boldsymbol{O}\{\varepsilon_J s_{\iota,J}\} \rangle_\alpha}$, which typically is an uninteresting constant value. The gauge transformation along each ferromagnetic chain leaves every $J_1$ interaction unchanged, maps the interaction $\tilde{J}_2$ into $\varepsilon_j \varepsilon_{j+1} \tilde{J}_2$ and leaves the disorder distribution invariant. The disorder average of any correlation between spins on different chains therefore vanishes identically, at every temperature. Namely, we have: $\overline{\langle s_{\iota,J} s_{\iota',J'} \rangle_\alpha} = 0$. We have to focus on the disorder average of the square of the pair correlation function of spins as follows.

$$\overline{\langle s_{\iota,J} s_{\iota',J'} \rangle_\alpha^2} = \int \sum_{<ij>} d\tilde{J}_2 \, P(\tilde{J}_2) \, \langle s_{i,j} s_{i',j'} \rangle_\alpha^2$$

(17)

where

$$\langle s_{i,j} s_{i',j'} \rangle_\alpha^2 = \frac{1}{\bar{Z}_\alpha^2} \mathrm{trace} \left\{ V^{j-1} s_i V^{j'-j} s_{i'} V^{m-j'+1} \right\}^2$$

(18)

following the procedure in [39], a statistical approach for the thermal average taken over all configurations of the model. The correlation between a pair of spins within the magnet equals to the trace of the product of the two spins and the transfer matrices of the model divided by the partition function. Here $V$ is a transfer matrix. To evaluate the trace, Kaufman and Onsager [39] used an approach of spinor analyses that employed to determine the partition function of the lattice also [38] to derive the explicit formulas for the correlation function between two spins in the different rows of the ferromagnetic 2D Ising model. Figure 5 in [39] illustrated correlations for small distances as functions of temperature in the ferromagnetic 2D Ising model. The correlation functions for finite distances for the ferromagnetic 2D Ising model are larger than zero at finite temperature. Lebowitz proved the non-negativity of the pair correlation function of spins and derived bounds on the decay of correlations between two widely separated sets of particles in terms of the decay of the pair correlation in ferromagnetic Ising spin systems [46]. The situation becomes much more complicated in the competing 2D Ising model. As described above, the averaging procedure consists of the thermal average and the disorder average. After the disorder average, the pair correlation function of spins varies in between 0 and 1. Namely, we have: $0 < \overline{\langle s_{i,j} s_{i',j'} \rangle_\alpha^2} < 1$.

The thermal average of the four-point correlation function of spins for a fixed replica α in the competing 2D Ising model can be defined as:

$$\langle s_{i,j} s_{i',j'} s_{u,v} s_{u',v'} \rangle_\alpha = \sum_{\{s_{i,j}\}} \left\{ \frac{s_{i,j} s_{i',j'} s_{u,v} s_{u',v'} \cdot exp[-\beta H_\alpha]}{\bar{Z}_\alpha} \right\} \tag{19}$$

The subscripts $(i,j)$, $(i',j')$, $(u,v)$ and $(u',v')$ denote spin sites in the system. The disorder average of the four-point correlation function of spins for a fixed replica α is expressed as:

$$\overline{\langle s_{i,j} s_{i',j'} s_{u,v} s_{u',v'} \rangle_\alpha} = \int \sum_{<ij>} d\tilde{J}_2 \, P(\tilde{J}_2) \, \langle s_{i,j} s_{i',j'} s_{u,v} s_{u',v'} \rangle_\alpha$$

(20)

In the competing 2D Ising model, upon the relative positions of four spin sites at the lattice, several types of the four-point correlators with different geometric figures exist: 1) a horizontal line **(—)**; 2) a vertical line **(|)**; 3) a slash or diagonal line **(/)**; 4) a triangular (∆) or perpendicular lines (⊥); 5) a quadrilateral, including parallel lines **(||)**, a plaquette **(□)**, a rectangle **(□),** a rhomb (◊), a parallelogram (▱), a trapezoid and an irregular quadrilateral. Squarcini and Tinti derived exact analytic results of several four-point correlation functions including parallel, rhomboidal and rectangular ones, for the ferromagnetic 2D Ising model exhibiting phase separation [64]. Figures 4-6 in [64] showed that the analytic results are in excellent agreement with Monte Carlo simulations, in which the parallel and rectangular ones possess the non-negativity, but the rhomboidal one not. It is thought that the randomly distributed interactions with different signs would make the situation much more complicated for the four-point correlation function of spins in the competing 2D Ising model. However, the disorder average restricts the four-point correlation function of spins to vary in range (0, 1), as the pair correlation function of spins does.

For the purpose of this work, the key is the existence of nonlinear terms of base matrices $P_j \equiv \Gamma_{2j-1}$ and $Q_j \equiv \Gamma_{2j}$ in all these four-point correlators, which corresponds to the existence of nontrivial topological structures as well as long-distance spin entanglements. The off-diagonal part arisen from the nonlinear terms dominates the four-point correlation function. The four-point correlation function of spins with different geometric figures have different nonlinear terms of Γ matrices in details upon the relative positions of spins at the lattice, for instance, how many terms of the Γ matrices, how to arrange them, and so on. But, the existence of the nonlinear terms of the Γ matrices is the most determining factor. It is understood that these types of the four-point correlation function have similar behaviors, specially, with the same asymptotic formula in the limit cases of infinite distance and temperature. Thus, in this work, we do not distinguish them in details, but only focus on their basic characters.

The topological contributions arisen by the nonlinear terms of the Γ matrices is a global effect that is a topological invariant. It does not depend sensitively on the signs of interactions, the distance between the spins and the geometric figures. As long as the nontrivial topological structures exist, a non-zero topological contribution to the four-point correlation function of spins exists. It is extremely hard to diagonalize the transfer matrices to derive the explicit formulas for the four-point correlation function of spins for the present model, because of 1) the existence of the nonlinear terms of the spin operators (and the nontrivial topological structures) and 2) the existence of randomly distributed interactions with different signs. One needs to introduce the time average, together with the application of Jordan algebra, to deal with the non-commutative problems of the spinors [80,82]. Furthermore, the topological Lorentz transformations should be employed to trivialize the nontrivial topological structures [80,82]. Unfortunately, the explicit formulas for the topological Lorentz transformations are unknown for the present model.

Luckily, it is enough to inspect a particular property of the system to derive a criterion for the four-point correlation function of spins. The four-point connected correlation function of spins, which equals to the difference between the four-point correlation function and the product of two pair correlation functions of spins [15,43,63], is described as:

$$\overline{\Delta_\alpha}\left(s_{i,j}s_{i',j'}s_{u,v}s_{u',v'}\right) = \overline{\langle s_{i,j}s_{i',j'}s_{u,v}s_{u',v'}\rangle_\alpha} - \overline{\langle s_{i,j}s_{i',j'}\rangle_\alpha \langle s_{u,v}s_{u',v'}\rangle_\alpha} > 0 \tag{21}$$

for a fixed replica α in the competing 2D Ising model with

$$\overline{\langle s_{i,j}s_{i',j'}\rangle_\alpha \langle s_{u,v}s_{u',v'}\rangle_\alpha} = \int \sum_{<ij>} d\tilde{J}_2\, P(\tilde{J}_2)\, \langle s_{i,j}s_{i',j'}\rangle_\alpha \langle s_{u,v}s_{u',v'}\rangle_\alpha \tag{22}$$

In a simple case that $(u,v) = (i,j)$ and $(u',v') = (i',j')$, the formula (21) can be reduced to:

$$\overline{\Delta'_\alpha}\left(s_{i,j}s_{i',j'}s_{u,v}s_{u',v'}\right) = \overline{\langle s_{i,j}s_{i',j'}s_{u,v}s_{u',v'}\rangle_\alpha} - \overline{\langle s_{i,j}s_{i',j'}\rangle_\alpha^2} > 0 \tag{23}$$

Note that Eqs. (21) and (23) are validated for all the permutations of subscripts $(i,j)$, $(i',j')$, $(u,v)$ and $(u',v')$ in the second term of the right hand of equation. The difference $\overline{\Delta_\alpha}$ in eq. (21) or $\overline{\Delta'_\alpha}$ in eq. (23) is larger than zero in the temperature range from zero ($T$ = 0) to infinite ($T \rightarrow \infty$), because of nonlinear terms of the Γ matrices, which is a pure topological contribution. It maintains the positivity as long as the nontrivial topological structures exist in the system, even in the thermodynamical limit ($m \rightarrow \infty$ and $l \rightarrow \infty$). Only as the system is exactly at infinite temperature ($T$ = ∞), every reduced interactions $K_1 \equiv J_1/(k_B T)$ and $\widetilde{K}_2 \equiv$ $\tilde{J}_2/(k_B T)$ goes to zero so that the nontrivial topological structures disappear and this differences $\overline{\Delta'_\alpha}$ and $\overline{\Delta_\alpha}$ become zero. □

**3.2 Equivalence between the four-point connected correlation function of spins and the pair correlation function of energy levels of the competing 2D Ising model $\mathbf{M}^{\mathbf{2D}}_{\mathbf{FI+SGI}}$**

**Theorem 2 (First Correlation Equivalence Theorem).** The four-point connected correlation function $\overline{\Delta}(s_{i,j}s_{i',j'}s_{u,v}s_{u',v'})$ of spins in the competing 2D Ising model $M^{2D}_{FI+SGI}$ is equivalent to the pair correlation function $\overline{\langle E_j E_{j+\Delta j}\rangle}$ of energy levels.

**Proof of Theorem 2.**

The thermal average of the pair correlation function of energy levels for a fixed replica α in the competing 2D Ising model can be defined as:

$$\langle E_j E_{j+\Delta j}\rangle_\alpha = \sum_{\{s_{i,j}\}} \left\{ \frac{E_j E_{j+\Delta j} \cdot exp[-\beta H_\alpha]}{\bar{Z}_\alpha} \right\} \tag{24}$$

where $\Delta j$ (= 1, 2,…, $\Delta J \to \infty$) denotes the number of the gaps between the energy levels $E_j$ and $E_{j+\Delta j}$. Analogues to the definition of the pair correlation function of spins (Eq. (2) in [39]), we obtain the pair correlation function of energy levels:

$$\langle E_j E_{j+\Delta j}\rangle_\alpha = \frac{1}{Z} trace\{E_j V^{\Delta j} E_{j+\Delta j} V^{m-\Delta j}\} \tag{25}$$

After diagonalization, every energy level $E_j$ corresponds to the product of two Γ matrices. It is assured that the pair correlation function of energy levels contains nonlinear terms of the Γ matrices, corresponding to the nontrivial topological structures in the system. The disorder average of the pair correlation function of energy levels for a fixed replica α is expressed as:

$$\overline{\langle E_J E_{J+\Delta_J}\rangle_\alpha} = \int \sum_{<ij>} d\tilde{J}_2\, P(\tilde{J}_2)\, \langle E_j E_{j+\Delta j}\rangle_\alpha$$

(26)

Tabar investigated quenched averaged correlation functions of the random magnets and showed that the quenched averaged connected correlation functions of local energy density field can be written in terms of ordinary scaling operators which can be constructed by the difference of energy operators in two different replicas [66].

Kadanaoff evaluated the 2n-spin correlation function for the 2D Ising model at $T = T_c$ for the special case in which all the spins lie along a straight line, separated by many lattice constants [37]. From Eqs. (4.1) and (4.3) in [37], we obtain the following relation between the pair correlation function of energy levels and the four-point correlation function of spins:

$$\langle E(R_1)E(R_2)\rangle \approx \frac{1}{C(|r_1 - r_{1'}|)C(|r_2 - r_{2'}|)} \times \frac{\langle s_1 s_{1'} s_2 s_{2'}\rangle - \langle s_1 s_{1'}\rangle\langle s_2 s_{2'}\rangle}{\langle s_1 s_{1'}\rangle\langle s_2 s_{2'}\rangle} > 0$$

(27)

Readers refer to [37] for the definitions of the parameters $C$ in Eq. (27) and we have used $s$ to replace σ in the Kadanaoff's formulas. This relation demonstrates the basic character of the pair correlation function of the energy levels, which is proportional to the four-point connected correlation function of spins. This relation can be generalized to be valid for the spins lying with different geometric figures and further for the competing 2D Ising model. Thus we have:

$$\overline{\langle E_J E_{J+\Delta_J}\rangle_\alpha} \sim \overline{\Delta}_\alpha\left(s_{i,j} s_{i',j'} s_{u,v} s_{u',v'}\right) > 0$$

(28)

for every replica α in the competing 2D Ising model. Scanning over all the replica (α = 1, 2,…, $R$) assures the validity of the equivalent relation (28) for the whole competing 2D Ising model. Namely, we have:

$$\overline{\langle E_J E_{J+\Delta_J}\rangle} \sim \overline{\Delta}\left(s_{i,j}s_{i',j'}s_{u,v}s_{u',v'}\right) > 0$$

(29)

Therefore, the pair correlation function of the energy levels is equivalent to the four-point connected correlation function of spins in the competing 2D Ising model, which is always larger than zero at temperature from $T$ = 0 to $T \rightarrow \infty$. Such an equivalence covers all the correspondences between the combinations of all four-spin configurations and all two energy levels (including $E_j \rightarrow \infty$ as $T \rightarrow \infty$), which is a global property of the system with a full coverage. □

**3.3 Equivalence between the pair correlation functions of energy levels of the competing 2D Ising model $\mathbf{M}^{\mathbf{2D}}_{\mathbf{FI+SGI}}$ and the pair correlation functions of the nontrivial zeros of the Dirichlet function $L(\mathbf{s}, \boldsymbol{\chi}_{\mathbf{k}})$**

**Theorem 3 (Second Correlation Equivalence Theorem).** The energy-energy correlation function $\overline{\langle E_J E_{J+\Delta_J}\rangle}$ of the competing 2D Ising model $M^{2D}_{FI+SGI}$ is equivalent to the pair correlation function $\langle L_s L_{s+\Delta s}\rangle$ between the nontrivial zeros of the Dirichlet function $L(s, \chi_k)$.

**Proof of Theorem 3.**

Montgomery [51] revealed that the pair correlation function of primes is the same as the pair correlation function of the energy eigenvalues of a random matrix for the GUE [17,68]. As revealed in [76], the competing 2D Ising model is a GUE with its transfer matrices and rotation matrices being the random matrices. Namely, the nontrivial zeros of the Dirichlet function is the spectrum of an operator, $\boldsymbol{R} = \frac{1}{2}\boldsymbol{I} + i\boldsymbol{H}$ with the self-adjoint operator $\boldsymbol{H}$ as the Hamiltonian of the competing 2D Ising model. The distribution of the energy eigenvalues $E_j$ in the competing 2D Ising model is equivalent to the distribution of the nontrivial zeros $L_s$ of the Dirichlet function. The pair correlation function $\overline{\langle E_j E_{j+\Delta j}\rangle}$ of the energy eigenvalues is equivalent to the pair correlation function $\langle L_s L_{s+\Delta s}\rangle$ of the nontrivial zeros of the Dirichlet function. Namely, we have the following relation:

$$\overline{\langle E_j E_{j+\Delta j}\rangle} \sim \langle L_s L_{s+\Delta s}\rangle > 0 \tag{30}$$

where $\Delta s$ (= 1, 2,…, $\Delta S \to \infty$) denotes the number of the intervals between the nontrivial zeros of the Dirichlet function. In a particular case $T \to \infty$ and $E_j \to \infty$, the relation (30) is valid also for $L_s \to \infty$. □

### 3.4 Equivalence between the pair correlation functions of the nontrivial zeros of the Dirichlet function $L(s, \chi_k)$ and primes

**Theorem 4 (Third Correlation Equivalence Theorem).** The pair correlation function $\langle L_s L_{s+\Delta s}\rangle$ between the nontrivial zeros of the Dirichlet function $L(s, \chi_k)$ is equivalent to the correlation function $\langle p_n p_{n+\Delta n}\rangle$ between a pair of primes $p_n$ and $p_{n+\Delta n} = p_n + q$ for every even $q$.

**Proof of Theorem 4.**

The correlation function between prime pairs $p_n$ and $p_{n+\Delta n}$ can be defined as

[32]:

$$\langle p_n p_{n+\Delta n}\rangle = \sum_{p_n, p_{n+\Delta n}} logp_n logp_{n+\Delta n}$$

(31)

Here $p_n$ denotes the $n$-th prime, while $p_{n+\Delta n} = p_n + q$ denotes the $(n + \Delta n)$-th prime with $n$ = 1, 2,…, $N \to \infty$ and $\Delta n$ = 1, 2,…, $\Delta N \to \infty$. $q$ (= 2, 4,…, $X \to \infty$) is an even number. Note that $q$ = 2 is the smallest even, with the character as normal as other even numbers.

The Riemann zeta function $\zeta(s)$ is defined via analytic continuation of Euler formula over the complex plane [58,59,69],

$$\zeta(s) = \sum_{n=1}^{\infty} \frac{1}{n^s} = \prod_{p=2,3,5,7\ldots} \frac{1}{\left(1 - \frac{1}{p^s}\right)}$$

(32)

It can be observed from this definition that the nontrivial zeros of the Riemann zeta function are inversely proportional to the logarithm of primes. The one-to-one correspondence between the nontrivial zeros of the Riemann zeta function and the primes can be generalized to be appropriate for the Dirichlet function. Thus, the distribution of the primes behaves as that of the nontrivial zeros of the Dirichlet function. Consequently, the pair correlation function $\langle L_s L_{s+\Delta s}\rangle$ for the nontrivial zeros of the Dirichlet function is essentially equivalent to the correlation formula $\langle p_n p_{n+\Delta n}\rangle$ of primes $p_n$ and $p_{n+\Delta n} = p_n + q$ for every even $q$ [24,40,51]. We have:

$$\langle L_s L_{s+\Delta s}\rangle \sim \langle p_n p_{n+\Delta n}\rangle > 0$$

(33)

The relation (33) is validated for all the values $s$ and $n$ and also for the case with $T \to \infty,\ E_j \to \infty,\ L_s \to \infty$ and $p_n \to \infty$. □

**3.5 Correlation function of a pair of primes**

**Theorem 5 (Positive Prime Correlation Theorem):** The pair correlation function $\langle p_n p_{n+\Delta n}\rangle$ of primes $p_n$ and $p_{n+\Delta n} = p_n + q$ for every even $q$ is larger than zero.

**Proof of Theorem 5.**

To investigate the correlation function between prime pairs, one usually denotes the von Mangoldt function $\Lambda$(n) [8,12,29,30]:

$$\Lambda(n') = \begin{cases} logp, & if\ \ n' = p^k\ \ for\ \ some\ \ prime\ \ p \\ 0 & otherwise \end{cases}$$

(34)

Note that $\Lambda(n')$ is defined for the condition $n' = p^k$, in conjunction with the Dirichlet function $L(s,\chi_k)$. The von Mangoldt function $\Lambda(n)$ can be formulated in the generating function [12,28,30]:

$$-\frac{\zeta'}{\zeta}(s) = \sum_{n\prime=1}^{\infty} \frac{\Lambda(n')}{n^s}$$

(35)

where poles of $\frac{\zeta'}{\zeta}(s)$ can be defined as the differential of the Riemann zeta function [13]. Of course, $\frac{\zeta'}{\zeta}(s)$ can be generalized to $\frac{L\prime(s,\chi_k)}{L(s,\chi_k)}$ [12,28,30]. Thus, the pair correlation function between of the nontrivial zeros of a Dirichlet function can be written as:

$$\langle L_s L_{s+\Delta s}\rangle = \sum_{n\prime=1}^{N\prime} \Lambda(n')\Lambda(n' + \mathrm{h}) = \langle p_n p_{n+\Delta n}\rangle$$

(36)

for a fixed even integer $h$. Once again we reach the equivalence between the pair correlation function $\langle L_s L_{s+\Delta s}\rangle$ of the nontrivial zeros of the Dirichlet function and the pair correlation function $\langle p_n p_{n+\Delta n}\rangle$ of primes.

Keating and Smith [41] established, via a heuristic Fourier inversion calculation, that the Hardy–Littlewood twin-prime conjecture is equivalent to an asymptotic formula for the two-point correlation function of Riemann zeros at a height $E$ on the critical line and that the reverse implication also holds for this equivalence. Assuming the Riemann hypothesis, they calculated the pair correlation function in the sum of the diagonal terms $R_2^{diag}(\epsilon)$ and the off-diagonal terms $R_2^{off}(\epsilon)$. Moreover, the off-diagonal part of the pair correlation function of the nontrivial zeros of a Dirichlet function is connected with the mean value of the product of two von Mangoldt functions in arithmetic progression [8]. Note that the off-diagonal part dominates the pair correlation function.

Summarizing briefly the results obtained above: The four-point connected correlation function $\overline{\Delta}\left(s_{i,j}s_{i',j'}s_{u,v}s_{u',v'}\right)$ of spins in the competing 2D Ising model is equivalent to the pair correlation function $\overline{\langle E_J E_{J+\Delta J}\rangle}$ of energy levels, the pair correlation function $\langle L_s L_{s+\Delta s}\rangle$ of the nontrivial zeros of the Dirichlet function and the pair correlation function $\langle p_n p_{n+\Delta n}\rangle$ of primes. The four-point connected correlation function of spins are all positive, which is a global property covering all the combinations of spin configurations and scanning all the replicas ($\alpha = 1, 2, \ldots, R$) in the system, so that the pair correlation functions of the energy levels, the nontrivial zeros of the Dirichlet function and the primes keep the positivity for all possible values, covering all the energy gap $\Delta j$ (= 1, 2,…, $\Delta J \to \infty$), all the zero interval $\Delta s$ (= 1, 2,…, $\Delta S \to \infty$) of the Dirichlet function, and all the prime interval $\Delta n$ = 1, 2,…, $\Delta N \to \infty$. For a particularly interesting case, all the conclusions in the present work are validated certainly for the limits of $T \to \infty$, $E_j \to \infty$, $L_s \to \infty$, and $p_n \to \infty$, correspondingly. Notice that the results are validated for every even number $q$ (= 2, 4,…, $X$), while the twin primes correspond to $\Delta n$ = 1 and $q$ = 2. The global property of the competing 2D Ising model certifies the closure on the property of the positivity for the pair correlation function of primes. Thus, we have proven:

$$\alpha(h) = \lim_{N\prime \to \infty} \frac{1}{N'} \sum_{n\prime=1}^{N\prime} \Lambda(n')\Lambda(n' + h) > 0 \tag{37}$$

for every fixed even integer $h$. □

We have demonstrated the following results: The pair correlation function $\langle p_n p_{n+\Delta n} \rangle$ of primes $p_n$ and $p_{n+\Delta n} = p_n + q$ for every even $q$ is larger than zero, and the number of such prime pairs diverges as $N'$ $\to \infty$. These results are required for full mathematical proofs.

## 4. Conclusion

In conclusion, we have proven five theorems: 1) The four-point connected correlation function of spins in the competing 2D Ising model $M^{2D}_{FI+SGI}$ is larger than zero in the temperature range from zero ($T$ = 0) to infinite ($T \to \infty$), because of nonlinear terms of operators (Positive Four-Point Spin Correlation Theorem). 2) The four-point connected correlation function of spins is equivalent to the pair correlation function $\overline{\langle E_J E_{J+\Delta_J} \rangle}$ of energy levels in the competing 2D Ising model $M^{2D}_{FI+SGI}$ (First Correlation Equivalence Theorem). 3) The pair correlation function $\overline{\langle E_J E_{J+\Delta_J} \rangle}$ of the energy levels in the competing 2D Ising model $M^{2D}_{FI+SGI}$ is equivalent to the pair correlation function $\langle L_s L_{s+\Delta s} \rangle$ of the nontrivial zeros of the Dirichlet function (including the Riemann zeta function) (Second Correlation Equivalence Theorem). 4) The pair correlation function $\langle L_s L_{s+\Delta s_l} \rangle_{av}$ between the nontrivial zeros of the Dirichlet function $L(s, \chi_k)$ is equivalent to the correlation function $\langle p_n p_{n+\Delta n} \rangle$ between a pair of primes $p_n$ and $p_{n+\Delta n} = p_n + q$ with $\Delta n = 1, 2, \dots, \infty$ for every even $q$ (Third Correlation Equivalence Theorem). 5) The pair correlation function $\langle p_n p_{n+\Delta n} \rangle$ of the primes $p_n$ and $p_{n+\Delta n} = p_n + q$ for every even $q$ is larger than zero (Positive Prime Correlation Theorem). This work offers a new insight on the twin primes conjecture and a path to upgrade it.

**Acknowledgements**

This work has been supported by the National Natural Science Foundation of China under grant numbers 52671259 and 52031014. The author is grateful to Fei Yang for understanding, encouragement, support and discussion. The author appreciates the referees of Physics Letters A for their valuable comments that initiate this revision.

**Declaration of competing interest:** The author declares no competing interests.